\documentstyle[12pt]{article}

\begin{document}
\setlength{\parskip}{5mm}

\title{The New Redshift Interpretation Affirmed }
\author{Robert V. Gentry
      \\The Orion Foundation
      \\P.O. Box 12067
      \\Knoxville, TN 37912
       }
\date{astro-ph/9810051}
\maketitle

\begin{abstract}
In late 1997 I published ({\it Mod. Phys. Lett. A} {\bf 12} (1997) 2919;
astro-ph/9806280) the discovery of the New Redshift Interpretation (NRI) of
the Hubble relation and the 2.7K CBR, which showed for the first time that
it was possible to explain these phenomena within the framework of a
universe governed by Einstein's static-spacetime general relativity instead
of the Friedmann-Lemaitre expanding-spacetime paradigm. Recently Carlip and
Scranton (astro-ph/9808021) claim to have found flaws in this discovery
based on the assumption that the NRI represents a static cosmological model
of the universe. This assumption is incorrect, and I show their
misunderstanding of this fundamental point is what led them to come to
erroneous conclusions about the NRI. I show the NRI very definitely
encompasses an expanding universe wherein galaxies are undergoing Doppler
recession according to the Hubble relation and, moreover, that---contrary to
Carlip and Scranton's claim---that the NRI does yield the correct form of
the Hubble magnitude-redshift relation. Lastly I note that Carlip and
Scranton signally fail to respond to the general relativistic results
wherein I show (gr-qc/9806061) that the universe is governed by Einstein
static-spacetime general relativity, and not the Friedmann-Lemaitre
expanding spacetime paradigm on which Big bang cosmology is critically
hinged, and also the most embarrassing fact that the F-L paradigm has always
involved gargantuan nonconservation-of-energy losses amounting to the mass
equivalent of about thirty million universes, each with a mass of 10$^{21}$ suns.
\end{abstract}

For almost seven decades cosmologists have assumed the universe is governed
by Friedmann-Lemaitre expanding-spacetime general relativity, and that both
the Hubble relation and the 2.7K CBR have their origin in redshifts due to
universal spacetime expansion. A widely accepted corollary of this belief
has been that no other explanation of the Hubble relation and the 2.7K CBR
is possible except that due to expansion redhifts.

Despite its long acceptance, this corollary was recently shown to be
incorrect when I reported the discovery \cite{1} of A New Redshift Interpretation
(NRI) of the Hubble relation and the 2.7K CBR based on the premise that the
universe is governed by Einstein's static-spacetime general relativity,
rather than Friedmann-Lemaitre expanding-spacetime general relativity. In
the NRI's Einstein framework the redshifts responsible for the Hubble
relation and the 2.7K CBR are a combination of relativistic Doppler and
gravitational effects rather than being attributed to Friedmann-Lemaitre
spacetime expansion. The discovery of the NRI naturally raised the question
of whether the universe is governed by the Einstein static-spacetime
paradigm, or by the Friedmann-Lemaitre expanding spacetime paradigm.

To answer this crucially important cosmological question I subsequently
compared the general relativistic predictions of both paradigms, and made a
second discovery---namely, that the results of several general relativistic
experiments provide proof that the universe is governed by Einstein's
static-spacetime general relativity, not Friedmann-Lemaitre expanding
spacetime general relativity \cite{2}. As of early October 1998 I am unaware of
any attempt to refute this second discovery.

On the other hand, in their recent e-print \cite{3}, Carlip and Scranton (C\&S)
have attacked the analysis supporting the first discovery \cite{1}. Their e-print
lists several factors which they claim demonstrate the NRI is a failure. I
now demonstrate their conclusion results from both misunderstanding the
NRI's results and by mixing them with unwarranted assumptions, which in turn
lead to presumed contradictions.

Their first big misunderstanding---which leads them to make several errors
in their evaluation---is their claim that the NRI is a ``new static
cosmological model.'' This is wrong on two counts: The NRI is not a
cosmological model and definitely does not represent a static universe. On
page 2921 of my NRI paper we read the following: ``...the NRI attempts to
account for the Hubble relation and the 2.7K CBR by using Doppler and
gravitational redshifts embedded in a universe governed by static-space-time
general relativity.'' This quote shows that at present the NRI is only an
interpretation or description of the structure of the universe. As yet I
have not presented it as a cosmological model.

And it is definitely not a static description because galaxies are moving
away from the Center according to the Hubble relation. Indeed, the NRI is
distinguished from all previous attempts to describe the universe in that it
describes galaxies that are experiencing Doppler expansion within the
framework of a universe governed by static-space-time general relativity.
Clearly, then, the NRI universe governed by static-space-time general
relativity cannot possibly represent a static cosmological model. Thus when
C\&S characterize the NRI as a static cosmological model in Sections 1, 3, 4
and 5, and use this erroneous assumption to claim the NRI fails several
different tests, they are in reality waging a war against a straw-man
argument of their own devising.

Their errors in Section 4 are specially egregious. Because of their
erroneous claim of the NRI being a static model, they incorrectly conclude
the luminosity will have only one factor of $(1+z)$ in the denominator, and
from that incorrect deduction, they then conclude that the NRI will give
predictions contrary to the Hubble diagram. Hence, they say, the NRI must be
a failure. However, if C\&S had carefully read the NRI paper they would have
noted there are two redshifts which are combined in Eq. (2) of my NRI paper
\cite{1}, one due to gravity and the other due to Doppler recession between source
and the receiver. Thus, in the NRI, the flux of radiation received at earth
from any distant galaxy is spread over a sphere (area = $4\pi r^2$), and is
diminished by one redshift factor at the point of emission. The energy of
each photon is decreased by $1+z$ because of this redshift, and a second
redshift occurs because Doppler recession causes the rate at which photons
arrive at earth to be diminished by the same factor. The net result is that
in the NRI, which is based on Einstein's static-spacetime general
relativity, the flux we expect to receive from any distant source of
luminosity $L$ is
\begin{equation}
{\rm flux}_{\rm NRI} = \frac{L}{4\pi r^2(1+z)^2} {\rm erg\cdot cm}^{-2}{\rm s}^{-1},
\end{equation}
which is the actually the expression cosmologists use to relate the flux and
redshift on the assumption that the universe is undergoing
Friedmann-Lemaitre spacetime expansion \cite{4}. And, following standard
astronomical practice \cite{4}, the foregoing expression enables us to define an
effective luminosity distance for the NRI framework as
\begin{equation}
d_L=r(1+z).
\end{equation}
Given that the above definition applies to the NRI, we can then substitute
it in the definition for the distance modulus, \medskip
\begin{equation}
m-M=5(\log d_L-1),
\end{equation}
to obtain
\begin{equation}
m-M=5[\log r(1+z)-1]
\end{equation}
as being applicable to the NRI. The expression for $m-M$ in terms of $z$ can
now be obtained by substitution of $r$ in terms of $z$ from Eq. (2) of the NRI
paper \cite{1}. As C\&S appropriately note, in the case for $z<1$, a good
approximation for NRI's Eq. (2) is $Hr/c\approx z/(1+z)$. In this case the
above expression becomes,
\begin{equation}
m-M=5[\log cz-\log H]-5,
\end{equation}
which is the simplified Hubble magnitude-redshift relation, minus the now
superfluous $\Omega_0$ term (see ref. \cite{4}, page 448).

Thus C\&S's dire prediction that the NRI's expression for $z$ is way out of
sync with the observational data is not only wrong, we find just the
opposite is true. The fact is that the above relation shows the NRI does
give the correct expression for the Hubble magnitude-redshift relation once
it is correctly interpreted in terms of galaxies undergoing Doppler
expansion combined with gravitational redshifts.

Likewise, also in Section 4---the section C\&S claim is of greatest
importance because it presumes to deal with observational data---their
assumption of the NRI as a static cosmological model also leads them to
erroneously ascribe their Eq.\ (23) to the NRI, whereas in fact this is not
the case. More specifically, in this instance C\&S adopt the assumption of
constant quasar density---an assumption that is neither stated nor implied
in the NRI paper---and from that proceed to apparently show how the redshift
distribution of quasars in the NRI compares poorly with that based on the
flat FLRW model. In actuality, all they did here was to again prove
that---as with everything else in life---if you make a wrong assumption, you
will surely come to a wrong conclusion.

Now, concerning their discussion in Section 3 of the NRI's outer hydrogen
shell, and their claim of its instability, rapid evaporation and temperature
decline, the fact is that I envision the outer luminous, hot hydrogen shell
as being a thin spherical shell of overlapping galaxies, with a thickness of
about one galactic diameter. A thin shell of overlapping galaxies
effectively resolves the opacity problem as well as questions of short-time
radiative cooling and gravitational instability.

Next, concerning their criticism in Section 3 of the constant density
assumption, this is at best ill-founded. Just as with the standard
cosmology, which they would hope to defend, they should easily have realized
the constant density assumption in the NRI is a relative
assumption---meaning it is assumed to be valid for the present epoch wherein
the observations are being made. Thus C\&S err when they claim that their
Eq. (9) is a problem for the NRI; it becomes a problem only by the
imposition of certain cosmological constraints on the NRI, constraints which
I do not accept and which are not a part of the NRI framework \cite{1}. In this
respect it is also worthy to note that C\&S apparently overlooked the fact
that in the NRI the main component of the total density is due to that of
the vacuum; so it is evident that a modern, infinitesimally slow decrease in
the ordinary density---as per their Eq. (8)---has virtually no effect on the
dynamics of galactic recession in the NRI\ framework.

Thus, the failures that C\&S describe do not represent what is in my NRI
paper. They represent instead C\&S's mistaken attempts to place my paper
into a mold of their own construction. Nowhere is this more evident than in
Section 1. There they identify the NRI with a static cosmological model as
the prime reason for concluding that the NRI is not consistent with general
relativity. Completely aside from their misidentifying the NRI as a static
cosmological model, it is ironic that they raise the issue of consistency
with general relativity because in gr-qc/9086061 I have already reported on
two matters of considerable importance considering this point.  

First, among other things, my analysis fully exposes one of the best kept
secrets of Big Bang cosmology---namely, that the Friedmann-Lemaitre
expanding spacetime paradigm has always necessitated gargantuan
nonconservation of radiation energy losses  the equivalent to thirty million
universes like our own, each composed of 10$^{21}$ suns. Some cosmologists
are aware of this; some aren't. But, to the best of my knowledge, none have
ever chosen to publish or publicize this most embarrassing fact. Thus, for
all practical purposes, only a tiny fraction of physicists in other fields
are aware that Big Bang's Friedmann-Lemaitre spacetime redshifts involve
huge and continuing nonconservation of energy losses. For some reason C\&S
were not at all inclined to increase that tiny fraction by making reference
to my e-print gr-qc/9806061, which details the specifics of this result. 

Next, when C\&S attempt to disprove the NRI by arguing that their Eq. (24)
represents the truth about $z$, $H$, and $r$, they do so using the implicit
assumption that the universe is formatted according to FLRW expanding
spacetime general relativity. (Earlier herein I showed the NRI does agree
with the Hubble magnitude-redshift relation.) The problem is that they were
aware that my e-print, gr-qc/9806061, documents experimental general
relativity results which I claim conclusively demonstrate that the universe
is formatted by Einstein's static-spacetime general relativity, and not FLRW
expanding-spacetime general relativity which is necessary for Big bang
cosmology. For some reason, however, in their highly critical evaluation of
the NRI, C\&S completely omit any discussion---or even an acknowledgment of
the existence---of this result.  

Now it has always been my understanding that when scientists undertake to
critique a colleague's results, they are under the highest obligation to
fairly consider all the evidence that bears on a controverted topic, even
when that evidence contradicts a position that has long been considered
unimpeachable. But in this instance C\&S signally avoided dealing with the
very experimental evidence \cite{2} that contradicts the fundamental basis of
their attempt to discredit the NRI, evidence which at the same shows that
the Big bang theory is fallacious, and that the NRI, or some version of it,
must be the correct description of the structure of the universe. I can
think of only two reasons why  they have thus far chosen to remain silent on
such a crucially important topic---namely, that the evidence I cite in favor
of the universe being governed by Einstein's static-spacetime general
relativity is truly unimpeachable.
Second, at the close of their Section 4, C\&S criticize the
NRI for not having a guiding principle to account for primordial
nucleosynthesis, in contrast to what they say is the standard cosmology's
successful prediction of light element abundances. But what C\&S don't say
is that their portrayal of a successful prediction is predicated on the
existence of spacetime expansion redshifts, which in turn is predicated on
our universe being governed by expanding-spacetime relativity. By omitting
mention of the overwhelming evidence that our universe is governed by
static-spacetime---and hence cannot possibly exhibit expansion
redshifts---C\&S have conveniently ignored the very information which
disproves their success story. 

Having said this, I wish to close on a positive note. C\&S emphasize that
the NRI is a finely-tuned description of the universe. I fully agree with
this emphasis. Indeed, how could it ever be otherwise? Surely a universe
that is so obviously fine tuned as ours, must necessarily require a
description that is equally fine tuned!


\begin{thebibliography}{0}

\bibitem{1}Robert V. Gentry, {\it Mod. Phys. Lett. A} {\bf 12} (1997) 2919; astro-ph/9806280.

\bibitem{2}Robert V. Gentry and David W. Gentry, gr-qc/9806061.

\bibitem{3}Steven Carlip and Ryan Scranton, astro-ph/9808021.

\bibitem{4}J. Silk, {\it The Big Bang}, pp. 447--448, W. H. Freeman \& Co., Revised
edition, 1989. 

\end{thebibliography}
\end{document}